\documentclass[conference]{IEEEtran}
\IEEEoverridecommandlockouts
\usepackage{cite}
\usepackage{amsmath,amssymb,amsfonts}
\usepackage{algorithmic}
\usepackage{algorithm}
\usepackage{graphicx}
\usepackage{booktabs}
\usepackage{threeparttable}
\usepackage{textcomp}
\usepackage[aboveskip=5pt,belowskip=-10pt]{caption}
\usepackage{xcolor}
\newtheorem{remark}{Remark}
\usepackage[colorlinks,
            linkcolor=black,
            anchorcolor=black,
            citecolor=black]{hyperref}

\def\BibTeX{{\rm B\kern-.05em{\sc i\kern-.025em b}\kern-.08em
    T\kern-.1667em\lower.7ex\hbox{E}\kern-.125emX}}
\begin{document}

\title{Cooperative ISAC Network for Off-Grid Imaging-based Low-Altitude Surveillance}

\author{\IEEEauthorblockN{Yixuan Huang${}^1$, Jie Yang${}^{2,3}$, Chao-Kai Wen${}^{4}$, Shuqiang Xia${}^{5,6}$, Xiao Li${}^1$, and Shi Jin${}^{1,3}$}
\IEEEauthorblockA{${}^1$National Mobile Communications Research Laboratory, Southeast University, Nanjing 210096, China \\
${}^2$Key Laboratory of Measurement and Control of Complex Systems of Engineering, Ministry of Education, \\
Southeast University, Nanjing 210096, China\\
${}^3$Frontiers Science Center for Mobile Information Communication and Security, Southeast University, Nanjing 210096, China\\
${}^4$Institute of Communications Engineering, National Sun Yat-sen University, Kaohsiung 80424, Taiwan\\
${}^5$ZTE Corporation, Shenzhen 518057, China\\
${}^6$State Key Laboratory of Mobile Network and Mobile Multimedia, Shenzhen 518057, China\\
Email: \{huangyx, yangjie, li\_xiao, jinshi\}@seu.edu.cn, chaokai.wen@mail.nsysu.edu.tw, xia.shuqiang@zte.com.cn
}
}

\newcommand{\bl}[1]{\textcolor{blue}{#1}}
\newcommand{\rl}[1]{\textcolor{red}{#1}}
\newcommand{\pl}[1]{\textcolor{purple}{#1}}

\maketitle

\begin{abstract}
The low-altitude economy has emerged as a critical focus for future economic development, emphasizing the urgent need for flight activity surveillance utilizing the existing sensing capabilities of mobile cellular networks.
Traditional monostatic or localization-based sensing methods, however, encounter challenges in fusing sensing results and matching channel parameters.
To address these challenges, we propose an innovative approach that directly draws the radio images of the low-altitude space, leveraging its inherent sparsity with compressed sensing (CS)-based algorithms and the cooperation of multiple base stations.
Furthermore, recognizing that unmanned aerial vehicles (UAVs) are randomly distributed in space, we introduce a physics-embedded learning method to overcome off-grid issues inherent in CS-based models.
Additionally, an online hard example mining method is incorporated into the design of the loss function, enabling the network to adaptively concentrate on the samples bearing significant discrepancy with the ground truth, thereby enhancing its ability to detect the rare UAVs within the expansive low-altitude space.
Simulation results demonstrate the effectiveness of the imaging-based low-altitude surveillance approach, with the proposed physics-embedded learning algorithm significantly outperforming traditional CS-based methods under off-grid conditions.

\end{abstract}

\section{Introduction}
The low-altitude economy has experienced rapid growth in recent years, with unmanned aerial vehicles (UAVs) anticipated to play key roles in various applications, including food delivery, traffic monitoring, and agricultural irrigation \cite{zheng2024random,wu2024vehicle}.
However, the exponential increase in the number of UAVs necessitates continuous and all-weather surveillance of low-altitude activities to ensure flight safety and facilitate UAV trajectory planning \cite{wu2024vehicle,jing2024isac}.
Given the limitations of visible light cameras and light detection and ranging devices under adverse lighting or weather conditions, integrated sensing and communication (ISAC) offers an alternative approach for low-altitude surveillance. ISAC achieves this by leveraging existing mobile cellular networks without requiring additional sensors or hardware equipment \cite{liu2024cooperative,li2023towards}.
 
UAV surveillance algorithms based on cellular networks can be categorized into two paradigms: active and passive target sensing \cite{wu2024vehicle}.
In the active paradigm, UAVs are treated as cooperative devices that establish communication links with base stations (BSs).
As a result, existing user localization algorithms can be utilized to determine UAV positions \cite{meles2023performance,huang2023joint}.
However, this approach is unable to monitor uncooperative or unauthorized UAVs, making it unsuitable for practical surveillance scenarios.

In contrast, the passive sensing paradigm addresses these limitations by enabling environmental sensing without the cooperation of the targets.
In monostatic radar sensing systems, ``range-angle'' maps of low-altitude space can be constructed through beamforming and scanning with large antenna arrays \cite{guan20213}.
However, fusing the sensing results from multiple BSs to improve performance requires complex decision-level fusion algorithms.
To fully exploit the sensing capabilities of mobile cellular networks, bistatic or multi-static sensing modes are preferred \cite{liu2024cooperative}.
In these modes, multiple BSs cooperate by transmitting and receiving sensing signals. Channel parameters are first estimated and then projected onto potential UAV locations using geometric relationships \cite{liu2024cooperative,huang2023joint}.
Nevertheless, this two-step method suffers from error propagation and challenges in matching channel parameters \cite{shi2024joint}.

In this study, we propose a novel approach that models the surveillance problem as a compressed sensing (CS)-based imaging problem \cite{dai2009subspace,tong2021joint}.
This approach directly uses raw channel state information (CSI) measurements for image formation, enabling seamless cooperative sensing among multiple BSs.
However, conventional CS models assume that targets are precisely located at predefined grid centers (i.e., on-grid taregts), which is unrealistic in low-altitude scenarios, as UAV trajectories are continuous and not confined to grid points (i.e., off-grid targets).
This mismatch can lead to modeling errors in the sensing matrix and distorted imaging results, as has been observed in channel estimation studies.
For instance, \cite{yang2012off} introduced a CS model that simultaneously estimates the sparse vector and its corresponding off-grid modeling errors.
However, this method relies on Taylor expansion, resulting in complex mathematical formulations for imaging problems.
Alternatively, \cite{li2024atomic} proposed using atomic norm minimization to optimize in gridless continuous space, but this method demands high computational resources, which may exceed the memory capacity of typical computers.
 
Inspired by recent advances in deep learning, \cite{huang2023off} employed a deep neural network (DNN) to map CSI measurements directly to sparse vectors, bypassing reliance on the sensing matrix.
Similarly, a primary result based on traditional on-grid CS methods was refined using a convolutional neural network (CNN) in \cite{wu2019deep}.
By embedding physical information, this approach can achieve high performance in sparse vector recovery \cite{guo2023physics}.
Following the principles in \cite{wu2019deep,guo2023physics}, we propose a physics-embedded off-grid imager for low-altitude surveillance.
Additionally, we adopt the online hard example mining (OHEM) method \cite{shrivastava2016training}, which adaptively focuses on samples that exhibit significant discrepancies from the ground truth, to design a novel loss function tailored for neural network training in this specific application.

The main contributions are summarized as follows:

\begin{itemize} 
\item We propose a cooperative sensing scheme for mobile cellular networks, leveraging CS-based imaging models and algorithms for low-altitude surveillance.

\item We design a physics-embedded off-grid imager and a novel loss function based on the OHEM principle, achieving high-performance low-altitude sensing.

\end{itemize}
 
The remainder of this paper is organized as follows:
Sec. II presents the system model.
Sec. III introduces the on-grid and off-grid imaging algorithms.
Simulation results and conclusions are provided in Sec. IV and Sec. V, respectively.

\section{System Model}
\label{sec-system-model}

\begin{figure}
    \centering
    \includegraphics[width=0.85\linewidth]{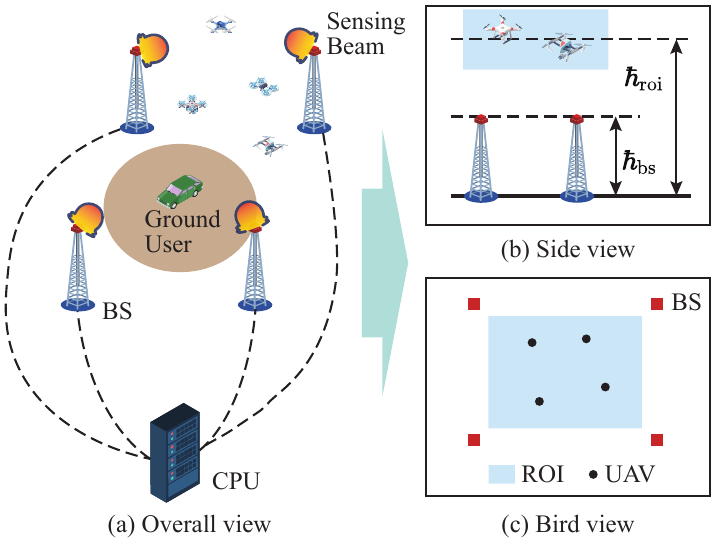} 
    \captionsetup{font=footnotesize}
    \caption{Illustration of the considered cooperative ISAC network.} 
    \label{fig-model}
\end{figure}

We consider an ISAC system operating within a 3D space represented as $[x,y,z]^{\text{T}} \in \mathbb{R}^3$, as illustrated in Fig. \ref{fig-model}.
The system comprises $N_{\text{b}}$ BSs located at the same altitude $\hbar_{\text{bs}}$, forming a convex region with $N_{\text{b}}$ edges in the horizontal plane.
Each BS is equipped with a full-duplex uniform planar array (UPA) consisting of $N_0\times N_0$ antennas.
The antenna arrays are oriented vertically to the ground, with their normal lines directed towards the center of the convex region.
The antenna spacing is $\lambda_0/2$, where $\lambda_0$ denotes the wavelength of the center carrier frequency $f_0$.
Self-interference at the full-duplex BSs is mitigated through antenna separation and specially designed beam directions \cite{zhang2015full}. Orthogonal frequency-division multiplexing (OFDM) signals are employed on $N_{\text{f}}$ subcarriers with a total bandwidth of $B$.

To focus on the sensing functionality of the cellular network in low-altitude spaces, we assume that the system operates in time-division modes, enabling both communication with ground users and flight activity surveillance.\footnote{ The BSs can also simultaneously provide communication services to ground users and monitor flight activities by transmitting separate beams toward the ground and the low-altitude space. Beamforming designs for such configurations are detailed in \cite{li2023towards}, but this is beyond the scope of this study.}
Each BS transmits sensing signals, which are subsequently received by all BSs, facilitating both monostatic and multi-static sensing modes.
The received signals are then sent to a central processing unit (CPU) for channel estimation and low-altitude image reconstruction.
Synchronization among the $N_{\text{b}}$ BSs is achieved via optical fibers.
The region of interest (ROI) is a large 3D space with an altitude of $\hbar_{\text{roi}}$, as illustrated in the blue region in Fig. \ref{fig-model}(b) and Fig. \ref{fig-model}(c).

\subsection{Signal Model}

During sensing symbol intervals, the transmitted signals from the $n_{\text{b1}}$-th BS form a wide beam to cover the ROI, as described in \cite{li2023towards}.
After scattering by targets within the ROI, the $n_{\text{s}}$-th signals received by the $n_{\text{b2}}$-th BS on the $n_{\text{f}}$-th subcarrier is expressed as 
\begin{equation}\label{eq-signal-model-1}
\begin{aligned}
\mathbf{s}_{n_{\text{b}1},n_{\text{b}2},n_{\text{f}},n_{\text{s}}}=&\ \sqrt{P_{\text{t}}}g\mathbf{F}^{\text{c}}_{n_{\text{b}2}}\mathbf{H}_{n_{\text{b}1},n_{\text{b}2},n_{\text{f}}}\mathbf{F}^{\text{p}}_{n_{\text{b}1}}\mathbf{x}_{n_{\text{b}1},n_{\text{f}},n_{\text{s}}}\\
&\ +\tilde{\mathbf{z}}_{n_{\text{b}1},n_{\text{b}2},n_{\text{f}},n_{\text{s}}},
\end{aligned}
\end{equation}
where $P_{\text{t}}$ denotes the transmit signal power, and $g$ is a constant related to the effective aperture size of the receiving antennas.
$\mathbf{H}_{n_{\text{b}1},n_{\text{b}2},n_{\text{f}}}\in \mathbb{C}^{N_0^2\times N_0^2}$ represents the channel from the $n_{\text{b1}}$-th BS, scattered within the ROI, and received by the $n_{\text{b2}}$-th BS.
$\mathbf{F}^{\text{p}}_{n_{\text{b}1}}$, $\mathbf{F}^{\text{c}}_{n_{\text{b}2}}$, and $\mathbf{x}_{n_{\text{b}1},n_{\text{f}},n_{\text{s}}}$ are the precoder, combiner, and transmitted data, respectively.
The designs of $\mathbf{F}^{\text{p}}_{n_{\text{b}1}}$ and $\mathbf{F}^{\text{c}}_{n_{\text{b}2}}$ are detailed in \cite{li2023towards}, while 
$\mathbf{x}_{n_{\text{b}1},n_{\text{f}},n_{\text{s}}}$ can be optimized  using specialized precoding techniques \cite{zheng2024random}.
Only signals passing through $\mathbf{H}_{n_{\text{b}1},n_{\text{b}2},n_{\text{f}}}$ are considered in \eqref{eq-signal-model-1}, owing to the sparsity of low-altitude space and tailored beamforming designs. Other multi-path channels are assumed to be included in the additive noise $\tilde{\mathbf{z}}_{n_{\text{b}1},n_{\text{b}2},n_{\text{f}},n_{\text{s}}}$.

Since UAV velocities are typically slow \cite{zheng2024random}, it is reasonable to assume that UAV locations and the channel $\mathbf{H}_{n_{\text{b}1},n_{\text{b}2},n_{\text{f}}}$ remains constant over $N_{\text{s}}=N_0^2$ symbol intervals.\footnote{
For example, assuming a subcarrier spacing of 60 kHz, a UAV velocity of 20 m/s, and $N_0=5$, the required coherent channel time is 0.4 ms, during which the UAV moves approximately 8 mm.}
By stacking the $N_{\text{s}}$ received signals, we obtain
\begin{equation}\label{eq-signal-model-2}
\begin{aligned}
\mathbf{S}_{n_{\text{b}1},n_{\text{b}2},n_{\text{f}}}=&\sqrt{P_{\text{t}}}\,g\mathbf{F}^{\text{c}}_{n_{\text{b}2}}\mathbf{H}_{n_{\text{b}1},n_{\text{b}2},n_{\text{f}}}\mathbf{F}^{\text{p}}_{n_{\text{b}1}}\mathbf{X}_{n_{\text{b}1},n_{\text{f}}}+\widetilde{\mathbf{Z}}_{n_{\text{b}1},n_{\text{b}2},n_{\text{f}}},
\end{aligned}
\end{equation}
where $\mathbf{S}_{n_{\text{b}1},n_{\text{b}2},n_{\text{f}}} = \big[\mathbf{s}_{n_{\text{b}1},n_{\text{b}2},n_{\text{f}},1}, \ldots, \mathbf{s}_{n_{\text{b}1},n_{\text{b}2},n_{\text{f}},N_{\text{s}}} \big]$. Similarly, $\mathbf{X}_{n_{\text{b}1},n_{\text{f}}}$ and $\widetilde{\mathbf{Z}}_{n_{\text{b}1},n_{\text{b}2},n_{\text{f}}}$ are defined as stacked versions of $\mathbf{x}_{n_{\text{b}1},n_{\text{f}},n_{\text{s}}}$ and $\tilde{\mathbf{z}}_{n_{\text{b}1},n_{\text{b}2},n_{\text{f}},n_{\text{s}}}$, respectively.

\subsection{Channel Model}

The channel model for $\mathbf{H}_{n_{\text{b}1},n_{\text{b}2},n_{\text{f}}}$ is defined as follows. The $(n_{\text{t}},n_{\text{r}})$-th element of $\mathbf{H}_{n_{\text{b}1},n_{\text{b}2},n_{\text{f}}}$ which represents the sensing channel from the $n_{\text{t}}$-th antenna of the $n_{\text{b1}}$-th BS to the $n_{\text{r}}$-th antenna of the $n_{\text{b2}}$-th BS on the $n_{\text{f}}$ subcarrier, can be given as \cite{huang2024fourier}
\begin{multline}\label{eq-continuous-sensing-channel-model} 
h^{\text{s}}_{n_{\text{t}},n_{\text{r}},n_{\text{f}}} = \iiint \frac{\breve{\sigma}(x, y, z)}{4 \pi d_{1}(x,y,z) d_{2}(x,y,z)} \\
\times e^{-j 2\pi\frac{\left(d_{1}(x,y,z)+d_{2}(x,y,z)\right)}{\lambda_{n_{\text{f}}}}} dxdydz, 
\end{multline}
where $\breve{\sigma}(x, y, z)$ is a continuous function representing the scattering coefficient of the point located at $[x, y, z]^{\text{T}}$.
$d_1(x,y,z)$ and $d_2(x,y,z)$ denote the distances from the point $[x, y, z]^{\text{T}}$ to the $n_{\text{t}}$-th transmitting antenna and the $n_{\text{r}}$-th receiving antenna, respectively.
$\lambda_{n_{\text{f}}}$ is the wavelength of the $n_{\text{f}}$-th subcarrier.

Given the high degrees of design freedom for $\mathbf{F}^{\text{p}}_{n_{\text{b}1}}$ and $\mathbf{F}^{\text{c}}_{n_{\text{b}2}}$, and the random properties of $\mathbf{X}_{n_{\text{b}1},n_{\text{b}2},n_{\text{f}}}$, these matrices can be assumed to be full-rank.
From \eqref{eq-signal-model-2}, the estimate of $\mathbf{H}_{n_{\text{b}1},n_{\text{b}2},n_{\text{f}}}$ can be computed as
\begin{equation}\label{eq-LS-channel-estimation}
\widehat{\mathbf{H}}_{n_{\text{b}1},n_{\text{b}2},n_{\text{f}}}\!=\!\frac{1}{\sqrt{P_{\text{t}}} \, g}\mathbf{F}^{\text{c}\ -1}_{n_{\text{b}2}}\mathbf{S}_{n_{\text{b}1},n_{\text{b}2},n_{\text{f}}}\mathbf{X}_{n_{\text{b}1},n_{\text{f}}}^{-1}\mathbf{F}^{\text{p}\ -1}_{n_{\text{b}1}}.
\end{equation}
In the subsequent sections, the CSI measurement, $\widehat{\mathbf{H}}_{n_{\text{b}1},n_{\text{b}2},n_{\text{f}}}$, is used to reconstruct low-altitude radio images.

\section{Learned Model-Driven Off-Grid Imager}

In this section, we first formulate the low-altitude surveillance problem and apply CS-based algorithms to perform imaging using on-grid models. Subsequently, we propose a physics-embedded learning algorithm designed to enable accurate imaging under off-grid conditions.

\subsection{CS-Based Problem Formulation}

\begin{figure}
    \centering
    \includegraphics[width=0.8\linewidth]{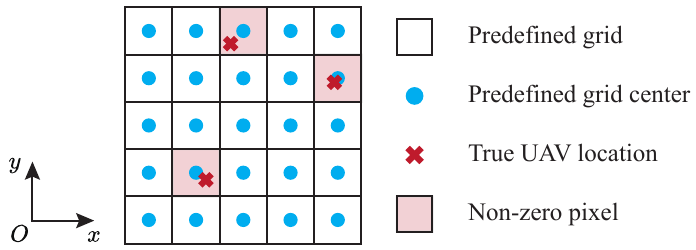}
    \captionsetup{font=footnotesize}
    \caption{A 2D illustration of low-altitude space discretization in $xOy$ plane.}
    \label{fig-discretization}
\end{figure}

In \eqref{eq-continuous-sensing-channel-model}, the low-altitude image is represented by the continuous function $\breve{\sigma}(x, y, z)$.
To reconstruct this image, the ROI is discretized into $N_{\text{v}} = N_{\text{x}}\times N_{\text{y}}\times N_{\text{z}}$ voxels, each with the size of $d_{\text{x}}\times d_{\text{y}}\times d_{\text{z}}$ ($\text{m}^3$).
Consequently, $\breve{\sigma}(x, y, z)$ is sampled into an $N_{\text{v}}$-dimensional vector $\boldsymbol{\sigma} = [\sigma_1, \ldots, \sigma_{N_{\text{v}}}]^{\text{T}}$, which represents the unknown image to be estimated.
A 2D slice of the low-altitude image in the $xOy$ is depicted in Fig. \ref{fig-discretization}.
UAVs are modeled as point targets randomly located within the ROI \cite{zhao2024buptcmcc}.  
The scattering coefficient of the $n_{\text{v}}$-th voxel is $\sigma_{n_{\text{v}}}$, corresponds to the scattering property of the UAV in that voxel (indicated by pink voxels in Fig. \ref{fig-discretization}). If no target exists in the voxel,
$\sigma_{n_{\text{v}}}=0$ (represented by white voxels in Fig. \ref{fig-discretization}). 
For simplicity, UAVs are assumed to be located at voxel centers in this section, serving as ``on-grid'' scatterers for modeling and analysis.
Errors introduced by the on-grid assumption are discussed in Sec. \ref{sec-algo-off-grid}.

Using the cascaded channel model, the discrete form of \eqref{eq-continuous-sensing-channel-model} can be expressed as \cite{huang2024fourier,tong2021joint}  
\begin{equation}\label{eq-channel-model} 
h^{\text{s}}_{n_{\text{t}},n_{\text{r}},n_{\text{f}}} =\sum_{n_{\text{v}}=1}^{N_{\text{v}}}h_{n_{\text{t}},n_{\text{v}},n_{\text{f}}}\sigma_{n_{\text{v}}} h_{n_{\text{r}},n_{\text{v}},n_{\text{f}}}=\mathbf{h}_{n_{\text{t}},n_{\text{f}}}^{\text{H}}\text{diag}(\boldsymbol{\sigma})\mathbf{h}_{n_{\text{r}},n_{\text{f}}}. 
\end{equation}
Here, $\mathbf{h}_{n_{\text{t}},n_{\text{f}}} = [h_{n_{\text{t}},1,n_{\text{f}}}, \ldots, h_{n_{\text{t}},N_{\text{v}},n_{\text{f}}}]^{\text{T}}$, and
\begin{equation}
h_{n_{\text{t}},n_{\text{v}},n_{\text{f}}}=\frac{e^{-j2\pi d_{n_{\text{t}},n_{\text{v}}}/\lambda_{n_{\text{f}}}}}{\sqrt{4\pi}d_{n_\text{t},n_{\text{v}}}},
\end{equation}
where $d_{n_{\text{t}},n_{\text{v}}}$ is the distance from the $n_{\text{t}}$-th transmitting antenna to the $n_{\text{v}}$-th voxel.
Similarly, $\mathbf{h}_{n_{\text{r}},n_{\text{f}}}$ and $h_{n_{\text{r}},n_{\text{v}},n_{\text{f}}}$ are defined.
The $(n_{\text{t}},n_{\text{r}})$-th element of $\widehat{\mathbf{H}}_{n_{\text{b}1},n_{\text{b}2},n_{\text{f}}}$ can then be written as 
\begin{equation}
\begin{aligned}
y_{n_{\text{t}},n_{\text{r}},n_{\text{f}}} &= h^{\text{s}}_{n_{\text{t}},n_{\text{r}},n_{\text{f}}} + z_{n_{\text{t}},n_{\text{r}},n_{\text{f}}} = \mathbf{a}^{\text{H}}_{n_{\text{t}},n_{\text{r}},n_{\text{f}}}\boldsymbol{\sigma} + z_{n_{\text{t}},n_{\text{r}},n_{\text{f}}},
\end{aligned}
\end{equation}
where $\mathbf{a}^{\text{H}}_{n_{\text{t}},n_{\text{r}},n_{\text{f}}}=\mathbf{h}_{n_{\text{t}},n_{\text{f}}}^{\text{H}}\text{diag}(\mathbf{h}_{n_{\text{r}},n_{\text{f}}})$, and $z_{n_{\text{t}},n_{\text{r}},n_{\text{f}}}$  is additive noise from channel estimation. 
By stacking measurements across all transmitting antennas, receiving antennas, and subcarriers, the measurements from the $n_{\text{b1}}$-th BS transmitter to the $n_{\text{b2}}$-th BS receiver are 
\begin{equation}\label{eq-measurement-model-1}
\mathbf{y}_{n_{\text{b}1},n_{\text{b}2}}=\mathbf{A}_{n_{\text{b}1},n_{\text{b}2}}\boldsymbol{\sigma} + \mathbf{z}_{n_{\text{b}1},n_{\text{b}2}},
\end{equation}
where $\mathbf{A}_{n_{\text{b}1},n_{\text{b}2}}\in\mathbb{C}^{N_{\text{f}}N_0^4\times N_{\text{v}}}$ with $(n_{\text{t}},n_{\text{r}},n_{\text{f}})$-th row $\mathbf{a}^{\text{H}}_{n_{\text{t}},n_{\text{r}},n_{\text{f}}}$.
According to channel reciprocity, the cellular network with $N_{\text{b}}$ BSs can deduce $N_{\text{b}}(N_{\text{b}}+1)/2$ groups of measurements similar to \eqref{eq-measurement-model-1}.
Stacking all equations across the cellular network yields
\begin{equation}\label{eq-measurement-model-2}
\mathbf{y}=\mathbf{A}\boldsymbol{\sigma} + \mathbf{z},
\end{equation}
where $\mathbf{A}\in\mathbb{C}^{N_{\text{f}}N_0^4N_{\text{b}}(N_{\text{b}}+1)/2\times N_{\text{v}}}$, and $\mathbf{z}$ is zero-mean additive Gaussian noise. 

To reconstruct $\boldsymbol{\sigma}$ from $\mathbf{y}$, we leverage the sparsity of $\boldsymbol{\sigma}$, reflecting the relatively small number of UAVs compared to the total voxels. Using CS theory and convex relaxation, the sensing problem is formulated as  \cite{dai2009subspace,tong2021joint}:
\begin{equation}\label{eq-CS-problem}
\text{(P1)} \ \ \hat{\boldsymbol{\sigma}} = \arg \min_{\boldsymbol{\sigma}} \|\boldsymbol{\sigma}\|_1, \ \ \text{s.t.} \ \|\mathbf{y}-\mathbf{A}\boldsymbol{\sigma}\|^2\le\varepsilon,
\end{equation}
where $\varepsilon$ is a small threshold ensuring reconstruction accuracy. 
The formulation of problem (P1) differs from traditional CS problems in the following ways:
\begin{itemize}
    \item The sensing matrix $\mathbf{A}$ may have a large condition number due to the high correlation among channels caused by the compact arrangement of antennas.
    \item The number of rows of $\mathbf{A}$, determined by $N_0$, $N_{\text{b}}$, and $N_{\text{f}}$, may exceed its number of columns $N_{\text{v}}$, depending on system configurations.
\end{itemize} 
The large number of measurements provides rich information about the low-altitude space, mitigating the uncertainty introduced by the high condition number of $\mathbf{A}$ and facilitating the recovery of $\boldsymbol{\sigma}$.

It is important to note that our objective is not to derive highly precise UAV locations.
Instead, we aim to detect non-zero voxels and their associated scattering coefficients, which provide sufficient information for intrusion detection and trajectory planning applications.
As such, the accuracy of UAV localization is predefined by the grid size, which should be selected based on the system's sensing capabilities and the specific requirements of the application.

\subsection{CS-Based Imaging Algorithms}
\label{sec-algo-on-grid}

Various algorithms can be applied to solve problem (P1) \cite{tong2021joint}.
Considering the trade-offs among estimation accuracy, computational complexity, and prior requirements, we select the subspace pursuit (SP) algorithm \cite{dai2009subspace} for this study.
The SP algorithm utilizes iterative refinement to enhance reconstruction accuracy.
 
Before describing the SP algorithm, we define the residual signal calculation function as
\begin{equation}
\mathbf{y}_{\text{res}}(\mathcal{S}) = \mathbf{y} - \mathbf{A}_{\mathcal{S}} \times f_{\text{LS}}(\mathbf{y}, \mathbf{A}_{\mathcal{S}}),
\end{equation}
where $\mathcal{S}$ represents the sparse signal support, and $\mathbf{A}_{\mathcal{S}}$ is the sub-matrix of $\mathbf{A}$ containing columns indexed by $\mathcal{S}$.
The function $f_{\text{LS}}(\mathbf{y}, \mathbf{A}_{\mathcal{S}})$ computes the estimates of the non-zero values in $\boldsymbol{\sigma}$ using the measurement $\mathbf{y}$, the sensing matrix $\mathbf{A}_{\mathcal{S}}$, and the least squares (LS) algorithm.
Additionally, $f_{{\text{sel}},K}(\mathbf{y}, \mathbf{A})$ selects the indices corresponding to the largest $K$ absolute values of $\mathbf{A}^{\text{H}}\mathbf{y}$, where $K$ is the number of non-zero values in $\boldsymbol{\sigma}$.

The SP algorithm begins by identifying an initial support 
$\mathcal{S}_0$ with $f_{{\text{sel}},K}(\mathbf{y}, \mathbf{A})$.
The residual signal for this support, $\mathbf{y}_{\text{res}}(\mathcal{S}_0)$ is then computed. In the $i$-th iteration, the algorithm performs the following steps:
\begin{enumerate}
    \item Expand the support to $\tilde{\mathcal{S}}_i$ by combining $\mathcal{S}_{i-1}$ and the indices obtained from $f_{{\text{sel}},K}(\mathbf{y}_{\text{res}}(\mathcal{S}_{i-1}), \mathbf{A})$.
    \item Update the support as $\mathcal{S}_i=f_{{\text{sel}},K}(\mathbf{y}, \mathbf{A}_{\tilde{\mathcal{S}}_i})$.
    \item Update the residual signal as $\mathbf{y}_{\text{res}}(\mathcal{S}_i)$.
\end{enumerate}
The algorithm stops iterating when the residual signal is smaller than the threshold $\varepsilon$ or when the support stabilizes.
Finally, the estimate of the non-zero values in $\boldsymbol{\sigma}$ are computed as $f_{\text{LS}}(\mathbf{y}, \mathbf{A}_{\mathcal{S}_{{\text{final}}}})$, where $\mathcal{S}_{{\text{final}}}$ denotes the final support.
The detailed steps of the SP algorithm are summarized in Algorithm \ref{ag1}.
Since the exact number of targets is unknown in advance, a prior-based sparsity estimate $K^\circ$ is used.

\begin{algorithm}[t]
\caption{SP algorithm \cite{dai2009subspace}.}
\label{ag1}
\begin{algorithmic}[1]
        \STATE $\mathbf{input:}$ Prior-based sparsity $K^\circ$, sensing matrix $\mathbf{A}$, and measurement vector $\mathbf{y}$.

        \STATE $\mathbf{initialize:}$ Calculate initial support $\mathcal{S}_0 = f_{{\text{sel}},K^\circ}(\mathbf{y}, \mathbf{A})$, derive the residual $\mathbf{y}_{\text{res}}(\mathcal{S}_0)$, and set $i=0$.

        \STATE  $\mathbf{while} \  \|\mathbf{y}_{\text{res}}(\mathcal{S}_i)\|_2>\varepsilon \ \text{or}\  \mathcal{S}_{i-1} \ne \mathcal{S}_{i} \ (i \ge 1) \ \mathbf{do}$

        \STATE \hspace{0.5cm} $i=i+1$.

        \STATE \hspace{0.5cm} Derive $\tilde{\mathcal{S}}_i=\cup (\mathcal{S}_{i-1}, f_{{\text{sel}},K^\circ}(\mathbf{y}_{\text{res}}(\mathcal{S}_{i-1}), \mathbf{A}))$.

        \STATE \hspace{0.5cm} Renew the support as $\mathcal{S}_i=f_{{\text{sel}},K^\circ}(\mathbf{y}, \mathbf{A}_{\tilde{\mathcal{S}}_i})$.

        \STATE \hspace{0.5cm} Update the residual as $\mathbf{y}_{\text{res}}(\mathcal{S}_i)$.

        \STATE $\mathbf{end\ while}$

        \STATE $\mathbf{output:}$ The estimated ROI image $\hat{\boldsymbol{\sigma}}$, whose element values are $f_{\text{LS}}(\mathbf{y}, \mathbf{A}_{\mathcal{S}_{i}})$ at support $\mathcal{S}_{i}$ and zero elsewhere.

\end{algorithmic}
\end{algorithm}

\subsection{Learned Physics-Embedded Off-Grid Imager}
\label{sec-algo-off-grid}

While the SP algorithm described in the previous subsection provides an effective approach for low-altitude imaging, it assumes that UAVs are located precisely at predefined voxel centers.
The sensing matrix $\mathbf{A}$ is constructed based on these on-grid points, and only targets aligned with these grids can be detected.
However, in practical scenarios, UAVs are randomly distributed, making it highly unlikely for their locations to coincide exactly with the predefined grid points, as illustrated in Fig. \ref{fig-discretization}.

As a result, the measurements $\mathbf{y}$ are generated from UAV locations that are off-grid.
When traditional CS-based algorithms are employed to reconstruct low-altitude images, a model mismatch occurs in $\mathbf{A}$, leading to inaccuracies.
This mismatch can significantly distort the imaging results, providing erroneous information about the locations and characteristics of low-altitude targets.
Such distortions are demonstrated in Sec. \ref{sec-simu-off-grid}. To overcome the challenges posed by off-grid targets, we propose a hybrid approach that integrates on-grid models with deep learning techniques.

\subsubsection{Algorithm Design}
The proposed method adopts a two-step approach, where a DNN is employed after initial processing based on physical models.

In the first step, although the CS-based model does not perfectly match the off-grid scenario, a preliminary result is derived from the measurement $\mathbf{y}$ and the sensing matrix $\mathbf{A}$ as
\begin{equation}
\boldsymbol{\sigma}_{\text{pri}} = \mathbf{A}^{\text{H}}\mathbf{y}.
\end{equation}
Unlike the SP algorithm, which applies thresholding to the estimated image and may inadvertently discard valuable information, this step projects data from the measurement domain to the image domain without enforcing strict sparsity constraints.
However, due to off-grid errors in $\mathbf{A}$, $\boldsymbol{\sigma}_{\text{pri}}$ often lacks explicit and accurate image information.

\begin{figure}
    \centering
    \includegraphics[width=0.88\linewidth]{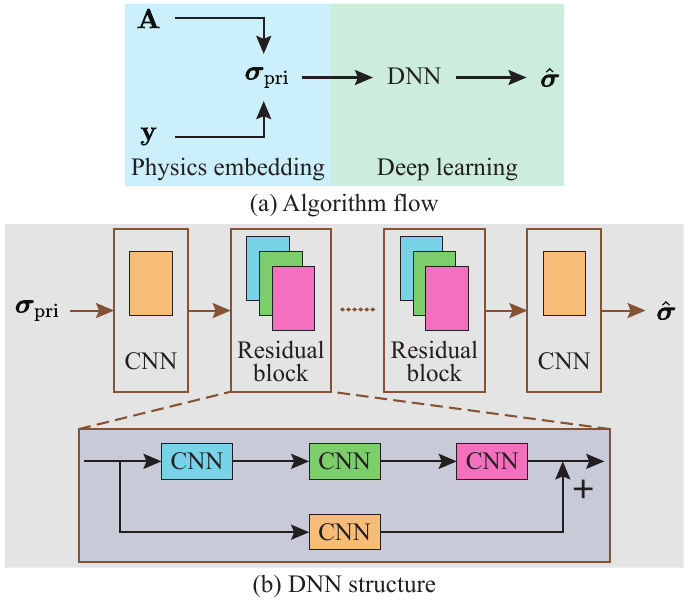} 
    \captionsetup{font=footnotesize}
    \caption{Illustration of the algorithm flow and the DNN structure.} 
    \label{fig-algorithm-flow}
\end{figure}

To refine the preliminary result $\boldsymbol{\sigma}_{\text{pri}}$,  it is fed into a DNN, which processes and outputs the final estimate $\hat{\boldsymbol{\sigma}}$, as illustrated in Fig. \ref{fig-algorithm-flow}. The DNN is designed with CNN layers and residual structures \cite{wu2019deep}, whose detailed parameters are stated in the simulation part. These residual structures facilitate faster network convergence and mitigate issues such as gradient vanishing or exploding during training.

In practical deployment, training data for the DNN can be collected using cooperative UAVs equipped with localization devices. This allows the model to learn effective mappings from preliminary CS results to accurate low-altitude images.

\subsubsection{Loss Function Design}
To refine $\boldsymbol{\sigma}_{\text{pri}}$ towards the ground truth image $\boldsymbol{\sigma}$, the mean square error (MSE) loss function has been utilized in previous studies \cite{huang2023off,wu2019deep}.
However, in the context of low-altitude imaging, this approach encounters significant challenges.
The sparsity of $\boldsymbol{\sigma}$  results in nearly all voxel values being zero, with only a few non-zero points.
Consequently, the DNN may converge to output all-zero images, achieving relatively low MSE values but failing to detect UAVs in the low-altitude space.

To address this issue and enhance both training performance and target detection rates, we adopt the OHEM method \cite{shrivastava2016training} to design an effective loss function.
OHEM was originally developed to mitigate class imbalance issues, guiding DNNs to focus on harder-to-detect targets with limited training samples. In this study, for each training image, we define:
\begin{itemize}
    \item {\bf Positive samples (hard samples):} Non-zero voxels, which represent UAV locations and occupy a small proportion of the total voxels.
    \item {\bf Negative samples (easy samples):} Zero voxels, which constitute the majority of the voxels.
\end{itemize}
For each predicted image: 
\begin{enumerate}  
    \item Compute and sum the MSEs of all positive samples to derive the loss $L_{\text{pos}}$.
    \item Compute the MSEs of all negative samples, sort them in the descending order, and select only the top $\eta K$ values to calculate $L_{\text{neg}}$, where $\eta$ is a hyper-parameter  that determines the number of selected negative samples.
\end{enumerate} 

The OHEM loss function for a single predicted image is then defined as
\begin{equation}\label{eq-loss-ohem}
L_{\text{ohem}} = \frac{L_{\text{pos}}+L_{\text{neg}}}{N_{\text{pos}}+N_{\text{neg}}},
\end{equation}
where $N_{\text{pos}}=K$  is the number of positive samples, and 
$N_{\text{neg}}=\eta K$ is the number of selected negative samples.
By adjusting  $\eta$, the influence of negative samples during training can be controlled. When $N_{\text{neg}}$ includes all negative samples, the loss function in \eqref{eq-loss-ohem} reduces to the traditional MSE loss, which tends to generate zero outputs.
Reducing $N_{\text{neg}}$ shifts the training focus toward detecting more non-zero voxels. Thus, $\eta$  should be carefully tuned to balance the training emphasis between positive and negative samples.

Considering the sparsity of low-altitude images (as formulated in \eqref{eq-CS-problem}), a sparse regularization term is incorporated into the loss function. The final loss function is expressed as
\begin{equation}\label{eq-loss-final}
L = L_{\text{ohem}}+\alpha \|\hat{\boldsymbol{\sigma}}\|_1,
\end{equation}
where $\alpha$ is a hyper-parameter controlling the weight of the regularization term.

\begin{remark}
The proposed learned physics-embedded imager for low-altitude surveillance offers several notable advantages:
\begin{enumerate}  
    \item Non-cooperative sensing: The method does not require cooperation from UAVs.
    \item Direct image formulation: CSI is directly used for image reconstruction, bypassing the need for delay and angular parameter estimation in localization-based methods, thus reducing error propagation.
    \item Efficient data fusion: Measurements from multiple BSs can be stacked and fused in their original forms, avoiding complex decision-level processing.
    \item Simultaneous multi-target detection: Multiple UAVs can be detected simultaneously, with computational complexity independent of the number of targets.
    \item High accuracy for off-grid targets: The imager maintains high sensing accuracy even for UAVs located off-grid.
\end{enumerate}

\end{remark}

\section{Numerical Results}

\subsection{Simulation Settings and Metrics}

We consider a mobile cellular network consisting of $N_{\text{b}}=4$ BSs,  which serve ground users while monitoring low-altitude flight activities.
The BSs are positioned at the corners of a square at a height of $\hbar_{\text{bs}}=20\ \text{m}$.
The center frequency is set to $f_0=2.6\ \text{GHz}$.
For simplicity, this study focuses on a 2D ROI at an altitude of  $\hbar_{\text{roi}}=40\ \text{m}$, with dimensions $120\ \text{m}\times 120\ \text{m}$.
This setup can be easily extended to a 3D ROI, which will be presented in our forthcoming journal paper.
The ROI is discretized into a $40 \times 40$ grid, with each pixel representing an area of $3 \,{\rm m} \times 3  \,{\rm m}$, sufficient for trajectory design and intrusion detection applications.

The additive noise power at each receiving antenna is set to -110 dBm \cite{li2023towards}.
The total noise power in the CSI measurements increases with the number of receiving antennas. 
The radar cross section (RCS) of UAVs is randomly generated according to a Gaussian distribution with a mean of 0.01 $\text{m}^2$ and the variance of 0.001 \cite{liu2024cooperative}.
The UAV scattering coefficient is calculated as the square root of its RCS \cite{huang2024fourier}.

Four metrics are employed to evaluate the sensing performance of the proposed algorithms:

\textbf{(1) MSE:} This metric quantifies the average per-pixel difference between the predicted image $\hat{\boldsymbol{\sigma}}$ and the ground truth $\boldsymbol{\sigma}$:
\begin{equation}\label{eq-mse}
\text{MSE}=\frac{1}{N_{\text{v}}}\|\hat{\boldsymbol{\sigma}}-\boldsymbol{\sigma}\|^2_2.
\end{equation}

\textbf{(2) Structural similarity index measure (SSIM):} This metric evaluates the structural similarity between the predicted image $\hat{\boldsymbol{\sigma}}$ and the ground truth $\boldsymbol{\sigma}$ \cite{wang2024dreamer}:
\begin{equation}\label{eq-ssim}
\text{SSIM}=\frac{\left(2 \mu_{\boldsymbol{\sigma}} \mu_{\hat{\boldsymbol{\sigma}}}+c_{1}\right)\left(2 \theta_{\boldsymbol{\sigma} \hat{\boldsymbol{\sigma}}}+c_{2}\right)}{\left(\mu_{\boldsymbol{\sigma}}^{2}+\mu_{\hat{\boldsymbol{\sigma}}}^{2}+c_{1}\right)\left(\theta_{\boldsymbol{\sigma}}^{2}+\theta_{\hat{\boldsymbol{\sigma}}}^{2}+c_{2}\right)},
\end{equation}
where $\mu_{\boldsymbol{\sigma}}$ ($\mu_{\hat{\boldsymbol{\sigma}}}$) and $\theta_{\boldsymbol{\sigma}}^{2}$ ($\theta_{\hat{\boldsymbol{\sigma}}}^{2}$) denote the mean and variance of are the average and variance of $\boldsymbol{\sigma}$ ($\hat{\boldsymbol{\sigma}}$), respectively.
$\theta_{\boldsymbol{\sigma} \hat{\boldsymbol{\sigma}}}$ is the covariance between $\boldsymbol{\sigma}$ and $\hat{\boldsymbol{\sigma}}$.
$c_{1}$ and $c_{2}$ are constants to stabilize the calculation.
SSIM ranges from 0 to 1, with higher values indicating greater similarity between $\boldsymbol{\sigma}$ and $\hat{\boldsymbol{\sigma}}$.

\textbf{(3) Detection rate (DR):} This metric represents the proportion of true targets in the ground truth image that are correctly identified in the reconstructed image.

\textbf{(4) False detection rate (FDR):} This metric indicates the proportion of targets in the reconstructed image that do not exist in the ground truth labels.

\subsection{Results and Discussions}

This subsection evaluates the sensing performance under various scenarios.
First, the SP algorithm is analyzed to assess the impact of system configurations on sensing performance under on-grid conditions.
Next, sensing results of different methods under off-grid conditions are compared.
Finally, the influence of the parameter $\eta$ in NN training is  discussed.

\subsubsection{Sensing Performance Evaluation of the SP Algorithm under On-Grid Conditions}

\begin{figure}
    \centering
    \includegraphics[width=0.85\linewidth]{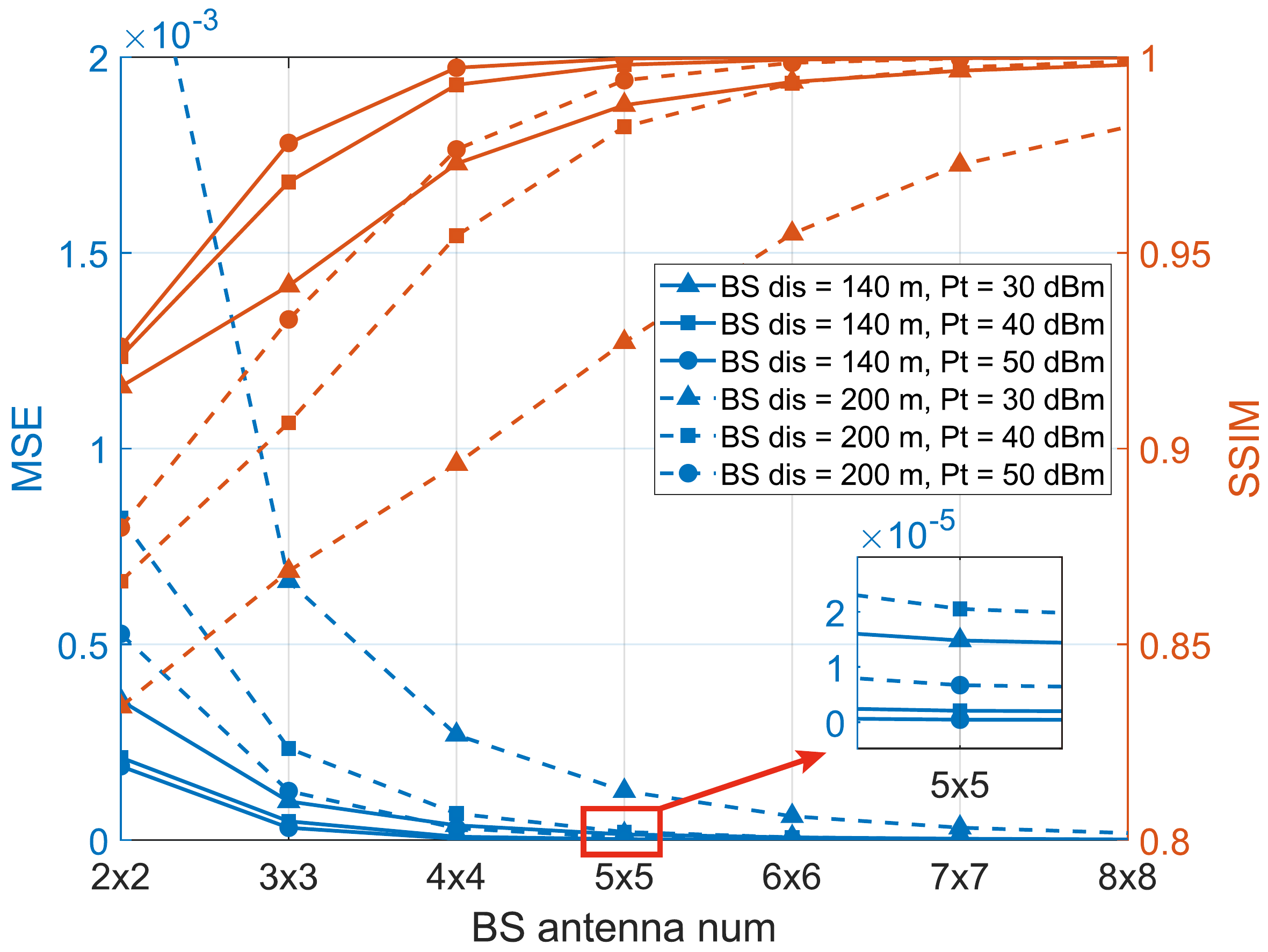} 
    \captionsetup{font=footnotesize}
    \caption{Sensing performance of the SP algorithm under on-grid conditions with varying system configurations.} 
    \label{fig-result1}
\end{figure}

We evaluate the sensing performance of the proposed imaging-based method using the MSE and SSIM metrics.
The results in Fig. \ref{fig-result1} represent the average of 1,000 Monte Carlo simulations.
As the number of BS antennas increases, more CSI measurements become available for reconstructing low-altitude images.
Consequently, the MSE decreases, and the SSIM improves, indicating enhanced sensing performance.
Specifically, with $5 \times 5$ UPAs at the BSs, nearly optimal sensing results are achieved when $P_{\text{t}}\ge 40 \ \text{dBm}$,  and the distance between BSs is 140 m.
Increasing the distances between BSs can degrade sensing performance due to higher channel correlations among antennas, which increase the condition number of the sensing matrix.
However, this negative impact can be partially mitigated by increasing the transmit power or employing larger transceiving antenna arrays.
These results demonstrate that low-altitude surveillance using CS-based imaging algorithms is effective when the system configurations are properly designed.

\subsubsection{Sensing Performance Comparison under Off-Grid Conditions}
\label{sec-simu-off-grid}

\begin{table}[t]
    \vspace{0.7cm}
    \renewcommand{\arraystretch}{1.3}
    \centering
    \fontsize{8}{8}\selectfont
    \captionsetup{font=small}
    \caption{Sensing performance comparison with different algorithms under off-grid conditions.}\label{tab-compare}
    \begin{threeparttable}
        \begin{tabular}{c|ccccc}
            \specialrule{1pt}{0pt}{-1pt}
            Methods & MSE & SSIM & DR & FDR \\
            \hline
            SP & 0.0067 & 0.6909 & 0.4652 & 0.6941 \\
            $\mathbf{A}^{\text{H}}\mathbf{y}$ & 0.0308 & 0.0534 & 0.4556 & 0.8599 \\
            DNN$\sphericalangle\mathbf{y}$ & 0.0037 & 0.6895 & 0 & 0 \\
            Model+DNN$\sphericalangle$SP & 0.0033 & 0.7778 & 0.3506 & 0.1528 \\
            Model+DNN$\sphericalangle\mathbf{A}^{\text{H}}\mathbf{y}$ & \textbf{0.0009} & \textbf{0.9186} & \textbf{0.8624} & \textbf{0.0329} \\
            \specialrule{1pt}{0pt}{-1pt}
        \end{tabular}
    \end{threeparttable}
    \vspace{-0.4cm}
\end{table}

\begin{figure}[t]
    \centering
    \includegraphics[width=0.9\linewidth]{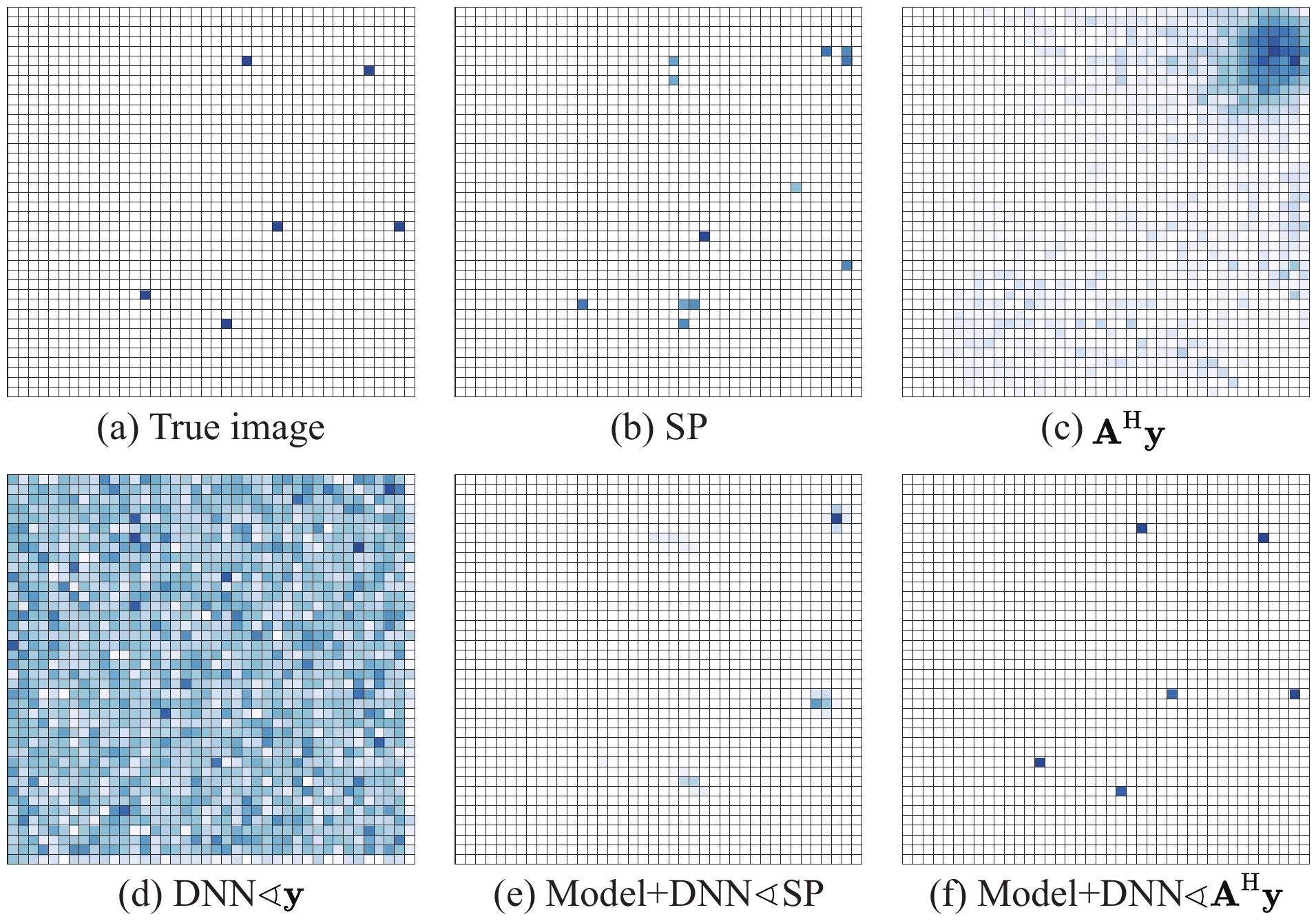}
    \captionsetup{font=footnotesize}
    \caption{Sensing results of various methods under off-grid conditions.} 
    \label{fig-result2}
\end{figure}

We evaluate the sensing performance of various imaging methods for off-grid UAV positions, including the SP algorithm with an on-grid sensing matrix, the intermediate sensing result $\mathbf{A}^{\text{H}}\mathbf{y}$, and a DNN trained using the CSI $\mathbf{y}$ as input (DNN$\sphericalangle\mathbf{y}$).
The proposed physics-embedded learning method, described in Sec. \ref{sec-algo-off-grid}, is validated with two configurations: training with the outputs of the SP algorithm (Model+DNN$\sphericalangle$SP) and training with $\mathbf{A}^{\text{H}}\mathbf{y}$ (Model+DNN$\sphericalangle\mathbf{A}^{\text{H}}\mathbf{y}$).

The experiments use a BS distance of 140 meters, a $5 \times 5$ antenna array, and a transmit power of $P_{\text{t}}=40 \ \text{dBm}$.
The DNN architecture includes six residual blocks with convolutional layers of 64, 128, 128, 128, 64, and 32 channels, respectively. The training process spans 200 epochs with an initial learning rate of 0.001. A total of 100,000 training images are randomly generated in MATLAB, with 10\% used for validation, and an additional 10,000 images are reserved for testing.

The simulation results in Table \ref{tab-compare} show that the proposed ``Model+DNN$\sphericalangle\mathbf{A}^{\text{H}}\mathbf{y}$'' method achieves the best performance across all metrics, significantly outperforming other methods.
Imaging examples in Fig. \ref{fig-result2} further illustrate its effectiveness. The SP algorithm struggles to accurately detect target voxels, producing multiple false positives due to model mismatch. Similarly, the intermediate result $\mathbf{A}^{\text{H}}\mathbf{y}$ and the DNN trained with $\mathbf{y}$
(DNN$\sphericalangle\mathbf{y}$) yield noisy images with many spurious points, failing to capture meaningful target information. In contrast, the ``Model+DNN$\sphericalangle$SP'' method detects some targets but lacks completeness. The proposed ``Model+DNN$\sphericalangle\mathbf{A}^{\text{H}}\mathbf{y}$'' method achieves near-perfect image reconstruction, accurately identifying the voxels containing UAVs.  

\subsubsection{Influences of the Selected Number of Passive Samples in NN Training}

\begin{figure}
    \centering
    \includegraphics[width=0.85\linewidth]{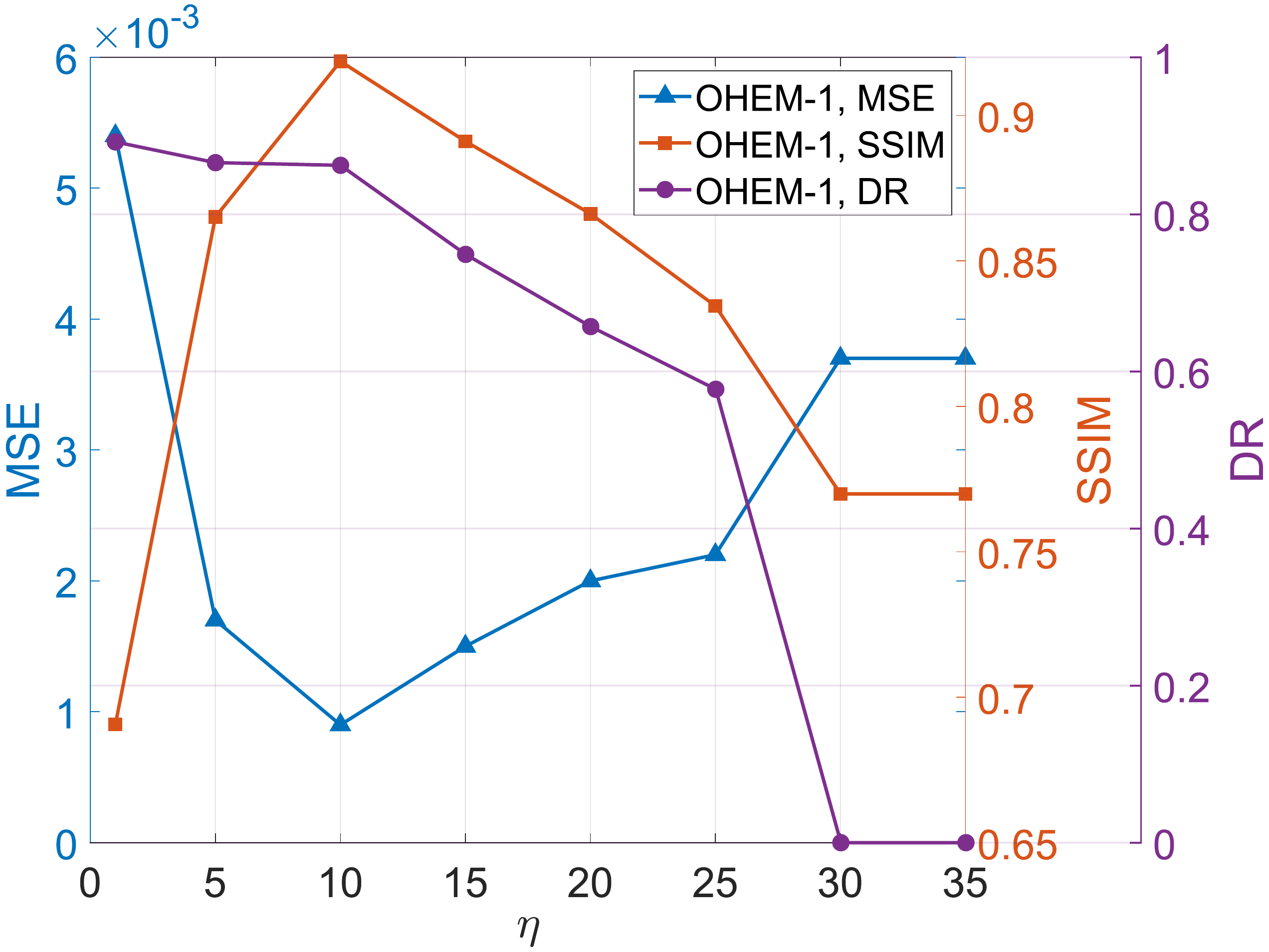} 
    \captionsetup{font=footnotesize}
    \caption{Sensing performance variation with different ratio $\eta$ of selected passive sample numbers to positive sample numbers.}
    \label{fig-result3}
\end{figure}

We analyze the impact of the parameter $\eta$ on sensing performance by training a series of NNs. The test dataset results for the trained NNs are presented in Fig. \ref{fig-result3}, showing distinct trends for the three metrics as $\eta$ varies.

When $\eta$ is small, the NN is biased toward generating a large number of non-zero pixels in the reconstructed images, which deviates from the true labels. As a result, the MSE is high, and the SSIM is low, indicating poor sensing performance. For instance, when $\eta=1$, the sensing performance is suboptimal, even though the DR reaches its peak.
As $\eta$ increases, the number of selected negative samples 
$N_{\text{neg}}$ grows, leading to improved focus on negative samples in the loss function.
At $\eta=10$, the simulation results demonstrate a balance between the attention on positive and negative samples. Consequently, the MSE reaches its minimum value, while the SSIM attains the highest value among all settings. Furthermore, the DR slightly decreases but remains close to its peak value, reflecting a balanced sensing performance.
When $\eta$ continues to grow beyond 10, the NN increasingly emphasizes generating images with more zero-value pixels. This shift causes the DR to decrease, accompanied by a deterioration in MSE and SSIM metrics. Finally, when $\eta \ge 30$, the NN outputs all-zero images, capturing no target information. These results underscore the critical role of $\eta$ in determining the sensing performance.

\section{Conclusion}
This study focuses on the issue of flight activity surveillance in the upcoming low-altitude economy era. The sensing function is achieved by utilizing existing cellular networks, and the sensing problem is innovatively modeled as a CS-based imaging problem. A physics-embedded learning method is proposed to address off-grid issues in the sensing matrix, and the OHEM principle is applied to design the loss function, enhancing NN training performance.
Simulation results show that imaging-based low-altitude surveillance can be realized through the cooperation of multiple BSs. The proposed physics-embedded learning method effectively retrieves images under off-grid conditions, significantly improving sensing performance compared to conventional CS-based algorithms.

\section{Acknowledgement}

This work was supported in part by the National Key Research and Development Program of China under Grant 2024YFE0200103;
in part by the National Natural Science Foundation of China (NSFC) under Grant 62261160576 and Grant 624B2036;
in part by the Key Technologies R\&D Program of Jiangsu (Prospective and Key Technologies for Industry) under Grant BE2023022-1 and Grant BE2023022;
in part by the Fundamental Research Funds for the Central Universities under Grant 2242023K5003.
The work of C.-K. Wen was supported in part by the National Science and Technology Council of Taiwan under the grant MOST 111-2221-E-110-020-MY3.

\bibliographystyle{IEEEtran}
\bibliography{ref}{}

\end{document}